\documentclass[prd,twocolumn,nofootinbib,floatfix,preprintnumbers]{revtex4-2}

\usepackage[utf8]{inputenc}
\usepackage{iftex}
\usepackage{amsmath}
\usepackage{amssymb}
\usepackage{color}
\usepackage{physics}
\usepackage{graphicx}
\graphicspath{
{./}
  {./media/}
}

\usepackage[caption=false]{subfig}
\usepackage{float}
\usepackage{dcolumn} 
\usepackage{bm}
\usepackage{soul}
\usepackage{hyperref}

\usepackage{comment}

\apptocmd{\sloppy}{\hbadness 10000\relax}{}{}

\ifPDFTeX
\else

\fi

\newcommand{\fO}{\mathcal{O}}

\newcommand{\lsigma}{l_{\sigma}}
\newcommand{\ltheta}{l_{\theta}}

\begin{document}

\title{Lattice study of the critical bubble in $\mathrm{SU(8)}$ deconfinement transition}

\preprint{HIP-2026-6/TH}

\author{Kari Rummukainen}\email{kari.rummukainen@helsinki.fi}
\affiliation	{Department of Physics and Helsinki Institute of Physics, \\
  		P.O. Box 64, FI-00014 University of Helsinki, Finland}	
  		
\author{Riikka Seppä}\email{riikka.seppa@helsinki.fi}
\affiliation	{Department of Physics and Helsinki Institute of Physics, \\
  		P.O. Box 64, FI-00014 University of Helsinki, Finland}	
  		
\author{David J. Weir}\email{david.weir@helsinki.fi}
\affiliation	{Department of Physics and Helsinki Institute of Physics, \\
  		P.O. Box 64, FI-00014 University of Helsinki, Finland}



\begin{abstract}
Strongly coupled theories are of phenomenological interest, for example as dark matter candidates.
Theories that can undergo first order thermal phase transitions are particularly appealing as potential sources of a stochastic gravitational wave background. Determining the expected gravitational wave signal from a first order phase transition requires accurate information on the bubble nucleation rate, but thus far for strongly coupled models these have relied on semiclassical methods. As a first step towards determining the nucleation rate, in this paper we study the confinement-deconfinement phase transition in a 4D SU(8) pure gauge model, using multicanonical Monte Carlo. Resolving the critical bubble for the first time in a pure Yang-Mills model, we determine the critical bubble probability and compare it to results from thin wall calculations. We also compare the effectiveness of different lattice pseudo-order parameters at resolving the condensation transition between the metastable phase and critical bubble branch, and point out the choice of order parameter is crucial to accurately resolve the critical configurations. 
\end{abstract}


\maketitle

\section{Introduction}

The possibility of a first order phase transition in the early universe is an exciting idea due to the potentially detectable gravitational wave signal such a transition would have produced (see Refs.~\cite{Hindmarsh:2020hop,Athron:2023xlk,Mazumdar:2018dfl} for recent reviews).
While no Standard Model transition is of first order\footnote{%
With the possible exception of the QCD transition at high quark chemical potential~\cite{Schmidt:2025ppy}. However, this case is not relevant for cosmology.}~\cite{Aoki:2006we,Bazavov:2011nk,Kajantie:1996mn,Csikor:1998eu}, many possible BSM transitions are.
Strongly coupled gauge field theories which exhibit a first order confinement phase transition are of particular interest, both as dark matter candidates (see e.g. Refs.~\cite{Huang:2020crf, LatticeStrongDynamics:2020jwi, Morgante:2022zvc, Gouttenoire:2023roe, Bruno:2024dha, Ayyar:2026wht}) and visible sector extensions~\cite{Fujikura:2023fbi}.

Because of the nonperturbative nature of these models near the transition, the input parameters needed for the gravitational wave power spectrum predictions are either determined from lattice simulations or estimated with semiclassical methods. A crucial quantity is the critical bubble nucleation rate, which can be estimated with the thin wall approximation (provided we know the surface tension and the latent heat) or from effective field theory descriptions~\cite{Huang:2020crf, Huber:2025qbl, Ayyar:2026wht}. While both of these methods can be calibrated with the results of lattice simulations, we note that in weakly-coupled gauge-scalar field theories, like the effective theory of the Standard Model, the bubble nucleation rate can also be computed directly from the lattice.

The lattice method, developed in Refs.~\cite{Moore:2000jw, Moore:2001vf} and more recently used in 
Refs.~\cite{Gould:2022ran, Gould:2024chm}, uses finite temperature multicanonical Monte Carlo simulations to compute the probability of a critical bubble forming in the metastable phase.
Critical bubble configurations are then evolved with real time simulations to obtain the dynamical prefactor of the bubble nucleation rate. A similar method has also been applied in sphaleron rate calculations, see Refs.~\cite{Moore:1998ge, Moore:1998swa, DOnofrio:2012phz, DOnofrio:2014rug, Annala:2025aci}.

In this work, we will show that even for a strongly coupled confining model, the bubble nucleation rate can be estimated straight from the lattice. 
Since real time simulations of strongly coupled models are not feasible, we will only be able to obtain the free energy of a critical bubble. This gives a rough upper limit to the nucleation rate, and helps shed light on the validity of the semiclassical approaches. Though earlier lattice studies of the critical bubble free energy in pure $\mathrm{SU}(N_c)$ exist~\cite{Kajantie:1992mh}, in these studies the bubbles were artificially inserted to the system, whereas with the present method the bubbles appear as a result of nucleation.

As a prototype for models with a confinement–deconfinement transition, 
we consider pure Yang-Mills theory. With $\mathrm{SU}(N_c)$ gauge group, the transition is of first order for $N_c \ge 3$, with the transition growing stronger with increasing $N_c$. We use $N_c = 8$, which, while being computationally more expensive than smaller $N_c$, has a stronger transition and shorter correlation length~\citep{Lucini:2005vg}. This enables larger levels of superheating, which in turn makes the critical bubbles smaller and thus easier to contain on a finite lattice. In this work we investigate only the transition from the confined to the deconfined phase. We anticipate this to be computationally simpler than the other direction, due to the fact that the expectation value of the fundamental order parameter, Polyakov loop, vanishes in the confined phase. We leave the confinement transition for future work. 

In a pure gauge deconfinement transition, the Polyakov loop usually plays the role of an order parameter. We, however, find evidence that without smearing or other improvements, the Polyakov loop struggles to resolve the critical bubbles. In practice, it cannot accurately identify lattice configurations which contain a critical bubble. 
We propose and test several improved order parameters, which enable us to resolve the critical bubble, to our knowledge for the first time in a confining model. This serves as a stepping stone towards determining the critical bubble free energy and hence estimating the nucleation rate from the lattice in physically more interesting confining models.

This work is structured as follows. In Sec.~\ref{sec:theory} we give a short overview 
of the lattice nucleation rate calculation.
Sec.~\ref{sec:lattice} describes the lattice formulation and simulation details. In Sec.~\ref{sec:results} we present our results, including visualizations of critical bubble configurations obtained from simulation, and comment on the different possible choices for an alternative order parameter. In Sec.~\ref{sec:compar} we compare our results to those obtained with the thin wall approximation.
We then conclude in Sec.~\ref{sec:disc}.

\section{Nucleation on the lattice} \label{sec:theory}

\subsection{Critical bubble free energy} \label{sect:nucleratecalc}

The nonperturbative method of determining the nucleation rate was first used in Refs.~\cite{Moore:2000jw, Moore:2001vf}, and has more recently seen use in e.g. Refs.~\cite{Gould:2022ran,Gould:2024chm}. The method is comprised of two parts. In the first part the statistical part of the nucleation rate is obtained via Monte Carlo sampling. The second part consists of obtaining the so-called dynamical prefactors of the rate by real time evolution. The real-time evolution in gauge-scalar models in Refs.~\cite{Moore:2000jw,Gould:2022ran} relies on weak gauge coupling which makes the gauge field infrared dynamics fully dissipative, and the real-time evolution can be done with Langevin-type updates~\cite{Bodeker:1998hm}.
In the strongly coupled regime this method is not applicable and real-time simulations are presently unfeasible.\footnote{The method presented by Barroso Mancha and Moore to measure the sphaleron rate in hot SU($N$) gauge theory~\cite{BarrosoMancha:2022mbj} can potentially be developed to obtain the real time critical bubble nucleation rate. However, this is beyond the scope of this article.}

In this work we thus restrict ourselves to the determination of the probabilistic part of the rate. It is obtained by calculating the probability of a critical bubble configuration with respect to the metastable phase, which in turn gives the critical bubble free energy. This can be thought of as the cost of having a critical-sized bubble of the stable phase in the surrounding metastable phase. 

\begin{figure}[tb]
   \centering
   \includegraphics[width=\linewidth]{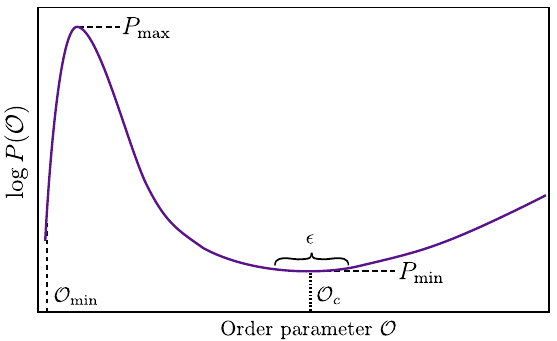} 
   \caption{Schematic of the probability distribution $P(\mathcal{O})$ of order parameter $\mathcal{O}$. Only the metastable phase and the mixed phase configurations separating the phases are shown. The order parameter value corresponding to the critical bubble is marked $\mathcal{O}_c$. Please see the main text for definitions of the other symbols.}
   \label{fig:schematic_probdistro}
\end{figure}

By `critical bubble' we mean a field configuration which is a saddlepoint of the free energy landscape. On a trajectory going from the metastable to the stable phase, the critical bubble is identified with the mixed-phase configuration with the maximal free energy, or the most suppressed of the configurations (see Fig.~\ref{fig:schematic_probdistro}). In the space of field configurations, the critical bubble lies on the transition surface, or separatrix,~\cite{Langer:1969bc, Hirvonen:2025hqn}, and is the most likely configuration of the transition surface. 
It then corresponds to a mixed-phase configuration which is equally likely to evolve into either phase, though in our case the time scale for this evolution is not clear.

Denoting the order parameter $\mathcal{O}$, we mark these configurations
as having an order parameter value $\mathcal{O}_c$.
These are then extremely suppressed with respect to the metastable phase. For the simulation parameters used in this study, there is  a suppression of at least a factor of $e^{-20}$, and usually more.

Since conventional Monte Carlo sampling would therefore be extremely inefficient, we employ multicanonical Monte Carlo~\cite{Berg:1992qua}, 
with automatic tuning as in Ref.~\cite{Moore:2000jw}. This means we sample configurations with a probability $p_{\mathrm{muca}} \propto \exp[-S + W(\mathcal{O})]$, where $W(\mathcal{O})$ is a weight function of the order parameter, iteratively constructed to make the sampled distribution of $\mathcal{O}$ approximately flat. 
The canonical distribution can then be obtained from the measured multicanonical distribution by reweighting,
\begin{equation}
P(\mathcal{O}) = e^{-W(\mathcal{O})} P_{\mathrm{muca}}(\mathcal{O}). \label{eq:canon_distro}
\end{equation}
Note that for efficient sampling the weight function should indeed result in a flat order parameter distribution (before re-weighting), but any (reasonable) weight function will yield
$P(\mathcal{O})$ upon reweighting, given enough statistics. In practice, it is not necessary to iterate $W(\mathcal{O})$ until we get a perfectly flat distribution $P_{\mathrm{muca}}(\mathcal{O})$, instead `an approximately flat' one will work, though the efficiency of the Monte Carlo sampling starts to degrade very rapidly with an inaccurate weight function.

The relative critical bubble probability with respect to the metastable phase, $P_c$, is given by the ratio of the sampled critical bubble configurations and the sampled metastable phase configurations. This can be obtained as the probability of having a configuration with an order parameter value $\mathcal{O}$ within an $\epsilon$-interval around the critical bubble order parameter value $\mathcal{O}_c$ (see Figure~\ref{fig:schematic_probdistro}), normalized by the probability of having a configuration belonging to the metastable phase,
\begin{align}
\tilde{P_c} &= \dfrac{ P(|\fO - \fO_c | < \epsilon/2 ) }{ \epsilon P(\fO < \fO_c) } \nonumber \\ 
&= \dfrac{1}{\epsilon } \int_{\fO_c - \frac{\epsilon}{2}}^{\fO_c + \frac{\epsilon}{2}} P(\fO) d\fO \left[ \int_{\fO_{\text{min}}}^{\fO_c} P(\fO) d\fO \right]^{-1}, \label{eq:integral_Pc}
\end{align}
where $\mathcal{O_\text{min}}$ is chosen so that the $P(\mathcal{O_{\text{min}}})$ is exponentially small and the metastable peak is included in the integration area, see Fig.~\ref{fig:schematic_probdistro}. 

This expression for the critical bubble probability is order parameter dependent, and has units of inverse order parameter.These effects would be cancelled out were we to compute the full nucleation rate, including the dynamical prefactors discussed earlier. To avoid the issue of a dimensionful expression, we consider an alternative formula.

The distribution $P(\mathcal{O})$ is in practice a histogram, with a minimum at $\mathcal{O}_c$, as the critical bubble corresponds to the most suppressed configuration of the transition, and a local maximum at some value $\mathcal{O}_{\text{ms}}$, corresponding to the metastable phase.
We write $P_\text{max} = P(\mathcal{O}_{\text{ms}})$ and  $P_\text{min} = P(\mathcal{O}_{c})$, as illustrated in Figure~\ref{fig:schematic_probdistro}. Then we can approximate
\begin{equation}
P_c = \dfrac{P_\text{min}}{P_\text{max}}. \label{eq:Pc_approx}
\end{equation}
This turns out to be a useful approximation because it always results in a dimensionless quantity, unlike Eq.~(\ref{eq:integral_Pc}). However, the critical bubble probability defined this way is still an order parameter-dependent quantity. The approximation holds well for high suppression and narrow peak. We shall use this method of calculating $P_c$ for the rest of the paper. 

From $P_c$ follows the free energy difference of a critical bubble~\cite{Moore:2001vf},
\begin{equation}
\dfrac{F_c}{T} = -\log P_c,
\end{equation}
which gives the statistical part of the nucleation rate $e^{-F_c/T}$. 
With Eq.~(\ref{eq:Pc_approx}), the free energy difference is thus given approximately by the log of the height ratio between the histogram metastable `bulk phase' peak and bubble minimum,
\begin{equation}
   \label{eq:alternative_Fc}
\dfrac{F_c}{T} = \log P_\text{max} - \log P_\text{min}.
\end{equation}
For the rest of this paper we shall use `bulk phase' to refer to the homogenous phase, that is, the non-mixed phase.

\begin{figure*}[tb]
   \centering 
   \includegraphics[width=\linewidth]{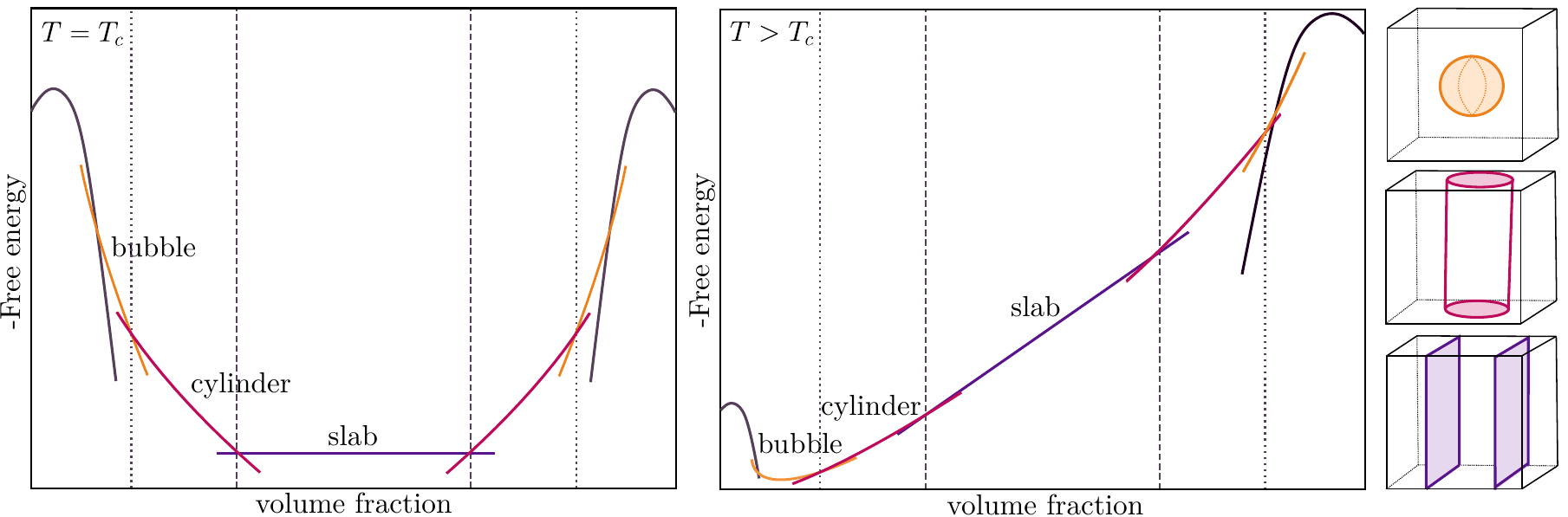} 
   \caption{Schematic illustration of different topological regimes predicted by the thin-wall approximation in a box with periodic boundaries. The left-hand figure is at the critical temperature, whereas the right-hand figure is above the critical temperature, where the stable phase peak is higher than the metastable phase peak. Regions where the different geometrical configurations are stable are separated by dashed and dotted lines. Dotted lines separate the bubble regime from the cylinder regime, and dashed lines the cylinder regime from the slab regime. The topologies can remain metastable beyond these lines, indicated by the free energy curves continuing beyond the dashed/dotted lines. The curves are shown with respect to the volume fraction of the deconfined, stable phase, which can be obtained from a normalized, intensive order parameter. 
The two peaks are schematic and represent the Gaussian bulk fluctuations in the two phases; for large lattices, these bulk phase fluctuations can be sufficiently broad as to cover the curve corresponding to the bubble branch.}
   \label{fig:thinw}
\end{figure*}

\subsection{Finite size effects}

Working with finite lattices with periodic boundary conditions, we encounter finite size effects. Most troublesome is trying to fit the bubble configurations on a lattice. If the lattice is too small the mixed-phase configurations which we implicitly assumed to be spherical in the earlier discussion will instead be cylindrical or slab-like.   

If we assume the walls are infinitesimally thin, to obtain a mixed-phase configuration with a spherical phase interface, the shortest spatial extent of the lattice needs to be at least triple the radius of the bubble.
This follows from considering which mixed-phase geometry minimizes the phase interface area, and thus the free energy, in a box of given size: when the volume of the bubble accounts for no more than $4\pi /81$ of the volume of the lattice, the spherical shape is favored over cylindrical, see Table 1 in Ref.~\cite{Moore:2000jw}. In Figure~\ref{fig:thinw} an illustration of regimes with different phase interface topologies is presented.

If the lattice is too small, the minimum of the probability (the maximum of the free energy) will correspond to non-spherical configurations. Thus to be able to obtain an accurate measurement of the critical bubble free energy, we will need to make sure that bubbles of the critical size (and a little larger) fit on the lattice. 

In reality, the different topological regimes are also separated by a barrier. 
A spherical configuration, for example, is metastable, i.e. stable with respect to a small fluctuation, until it is so large that is touches itself through the periodic boundary~\cite{Moore:2000jw}. There is a cost associated with deforming from a spherical mixed-phase configuration to a cylinder, for example. This means that if one wants to only sample spherical mixed-phase configurations, the volume fraction bound of $4\pi/81$ for a spherical mixed-phase geometry may not be strict. 

In a similar vein, there is also a condensation transition when local
fluctuations about the bulk phase condense into a spherical mixed-phase configuration~\cite{binder1980critical, Hallfors:2025key}. This degrades the efficiency of the Monte Carlo sampling, but luckily does not seem to be a significant problem for the volumes we simulate (although distinguishing between configurations on either side of the condensation transition remains a challenge, as we discuss below). There are also possible improvements to the multicanonical algorithm which would mitigate this issue~\cite{Hallfors:2025key}.

Fitting the critical bubble configurations on the lattice can be done in two ways: increasing the lattice size, or moving further away from the critical temperature $T_c$, in which case the critical bubble shrinks with respect to a fixed spatial volume. In the case of the deconfinement transition, this implies increased superheating. See Figure~\ref{fig:thinw} for an illustration of how the superheating changes the probabilities for the regimes with different phase interface topologies. 

However, increasing the spatial volume means increased computational costs, and more importantly makes the order parameter less reliable at identifying the bubble configurations. This can be seen by considering how bulk fluctuations and the critical bubble behave with volume. At constant temperature, the critical bubble has a fixed spatial size, so the volume fraction taken up by the bubble will decrease as $1/V$ with increasing spatial volume. On the other hand, the width of the bulk fluctuations decreases with increasing spatial volume as $1/\sqrt{V}$ \cite{Moore:2001vf}.

 Thus as volume $V$ increases, the bulk fluctuations will drown out the signal of the critical bubble. In essence, at some point, the order parameter will not assign different values to configurations with a bubble and those with large fluctuations, and thus the order parameter value can no longer be used to discriminate between bubble and bulk configurations.

In the right-hand panel of Fig.~\ref{fig:thinw}, this would mean that the bubble branch and its minimum would be covered by the left-hand Gaussian representing the bulk fluctuation in the metastable phase.

Thus we will need to find a balance between superheating and lattice size, or improve the performance of the order parameter. We will do both of these, with improvements to the order parameter discussed in Section~\ref{sec:Orderparam}.

\section{Lattice simulations} \label{sec:lattice}

\subsection{Simulation setup}

We use the standard Wilson action for $\mathrm{SU}(8)$,
\begin{equation}
   S = \beta \sum_{x, \mu > \nu} \left[  1 - \frac{1}{8}\Re \Tr U_{\mu\nu}(x) \right],
\end{equation}
where $\beta=2\,N_c/g^2$ is the inverse gauge coupling, and $U_{\mu \nu}(x)$ refers to the plaquette variable 
\begin{equation}
U_{\mu\nu}(x) = U_{\mu}(x)\,U_{\nu}(x+\hat{\mu})\,U^{\dagger}_{\mu}(x+\hat{\nu})\,U^{\dagger}_{\nu}(x) .
\end{equation}

The lattice is periodic with $L_s^3$ spatial extent, and $L_t$ temporal extent. With lattice spacing $a=a(\beta)$ we write these as $L_s = a N_s$, and $L_t = a N_t$, where $N_t$ and $N_s$ correspond to the number of lattice sites in the temporal and spatial directions, respectively. We will use $V = N_s^3$ in the following to refer to the lattice volume, by which we mean the total number of spatial lattice sites.

The time extent is related to the temperature through 
\begin{equation}
T = \dfrac{1}{N_t a},
\end{equation}
and thus varying the inverse lattice coupling $\beta$ at fixed time extent $N_t$ varies the temperature.

We simulate with only one choice of $N_t$, namely $N_t = 6$, but vary the spatial length from $N_s = 60$ to  $N_s = 80$. We study the system at three couplings $\beta_1 = 44.742$ and $\beta_2 = 44.712$, $\beta_3 = 44.682$. With $\beta_c = 44.562$~\cite{surfacetension}, these correspond to superheating of $\Delta \beta = \beta_c - \beta = -0.12, -0.15$ and $-0.18$, respectively.

The simulations are multicanonical \cite{Berg:1992qua, Moore:2000jw}. We use a standard mix of heatbath and overrelaxation updates, with 5 to 6 overrelaxation updates for each heatbath update, and a multicanonical accept-reject step after having updated all even or odd time-direction links (since only these affect the order parameter value when it is derived from the Polyakov loop). The simulation code uses the \texttt{hila} lattice field theory framework~\cite{hilacode}. 

Lattice simulations of SU($N$) gauge theory simulations at large-$N$ can suffer from topological freezing, i.e. the topological number evolves very slowly during the simulation. This freezing was observed in SU(8) phase transition simulations in Ref.~\cite{Rindlisbacher:2025dqw}. In this work the latent heat and the surface tension were measured to high accuracy. The effect of the topological freezing was seen to be very small, below the statistical accuracy of the results.  In the present study the statistical errors are much larger and hence the effect of the topological freezing is expected to be negligible.

In Table~\ref{tab:param} the different sets of simulation parameters are given. The set of parameters and the choice of order parameter affect the simulations from the beginning: the multicanonical weight function needs to be generated separately for each choice, though previous weight function results can be used to make educated starting guesses, which cut down on simulation time. 

A weight function is iteratively computed by tuning the weight to produce a quasi-flat order parameter distributions within a given order parameter range. The weight function iteration is ready when the updates cease to change the weight function significantly, and the resulting (non-reweighted) order parameter distribution is approximately flat.
After the weight functions have been obtained, we perform a multicanonical simulation with each set of simulation parameters until the system has tunneled several tens of times back and forth in a restricted sampling range.

We simulate the system only on a fixed interval of order parameter values, since it is not necessary to sample the whole range from metastable to stable bulk peak to obtain the critical bubble probability with respect to the metastable phase. In practice we set the range to include the metastable phase bulk peak and most of the bubble branch, so that the critical bubble minimum is included. In the multicanonical algorithm this is enforced by setting the weight $W(\mathcal{O})$ very large outside the desired sampling range, so that updates that take the ensemble out of range are always rejected. Our chosen range is sufficient to obtain the critical bubble free energy~\cite{Moore:2000jw}.

\begin{table}
\centering

\begin{ruledtabular}
\begin{tabular}{	c |	c |	c |	c |	c 	}
$N_t$ & $\beta$ & $\mathcal{O}$ & $n_{\text{smear}}$ & $N_s$ \\
\hline
6 & 44.682 & $\ltheta$ & 48 & 72, 80 \\ 
\hline
6 & 44.712 & $\ltheta$ & 48 & 60, 64, 80\\ 
 && $\lsigma$ & 48 & 60 \\
\hline
6 & 44.742 & $\ltheta$ & 48 & 60, 64, 72, 80\\ 
& & $\ltheta$ & 24 & 60 \\
& & $\lsigma$ & 48 & 60, 64, 72\\
& & $|l_p|$ & 0 & 60\\
\end{tabular}
\end{ruledtabular}

\caption{List of simulation parameters. Listed are the time extent, which is $N_t = 6$ for all runs, the value of $\beta$, which in turn determines the temperature, the order parameter $\mathcal{O}$ used for the multicanonical algorithm, the number of nearest-neighbour smearing steps $n_{\textrm{smear}}$ used in measurement of the order parameter $\mathcal{O}$, and the different spatial extents used. The different order parameters are defined in Eqs.~(\ref{eq:old_order_parameter}), (\ref{eq:theta_order_parameter}) and~(\ref{eq:sigma_order_parameter}).}
\label{tab:param}
\end{table}

\subsection{Order parameter}	\label{sec:Orderparam}

When considering a phase transition, a measurable whose value can be used to distinguish the two phases is needed. For our purposes, the measurable will need to distinguish the intermediate mixed phase configurations as well. For
the deconfinement-confinement transition, the volume averaged Polyakov loop is the usual choice for an order parameter. 

The Polyakov loop is defined on the lattice as
\begin{equation}
l_p(\vec{x}) = \Tr \prod^{N_t -1}_{t = 0} U_4(\vec{x}, t) \label{eq:Polyakov_loop}.
\end{equation}
The expectation value of the Polyakov loop vanishes in the confined phase, and is non-zero in the deconfined phase, where its value differs between the eight degenerate vacua by complex phases.
For our purposes, the choice of the vacuum does not matter, so we take the absolute value of the volume average as
\begin{equation}
|\bar{l}_p| = \left\lvert \dfrac{1}{N_s^3} \sum_{\vec{x}}  l_p(\vec{x})\right\rvert. \label{eq:old_order_parameter}
\end{equation}	

In the case of the critical bubble probability, some additional considerations are necessary.
At our volumes, it turns out that the conventional order parameter $|\bar{l}_p|$ cannot discriminate between bubble and bulk configurations. 

The issue stems from the width of the bulk phase fluctuations. We must thus find an alternative quantity which is less sensitive to the fluctuations.
One approach is to square the local order parameter before volume averaging~\cite{Moore:2000jw, Moore:2001vf}. In our case this is not entirely trivial, since the Polyakov loop is a complex variable for $\mathrm{SU}(8)$, and the volume averaging before taking the absolute value or squaring is necessary to cancel noise in the complex and real parts.

To reduce the noise before squaring the Polyakov loop field, we use iterative nearest-neighbour smearing. We perform $n$ consecutive weighted averages over nearest neighbours,
\begin{equation}
l^{(m)}_{s}(\vec{x}) = \dfrac{1}{4} \left( l^{(m-1)}_{s}(\vec{x}) + \dfrac{1}{2} \sum_{\hat{i}} l^{(m-1)}_{s}(\vec{x} + \hat{i}) \right), \label{eq:smeared_loop_field}
\end{equation}
for $m = 1, \ldots, n$ and where $l^{(0)}_{s} = l_p$, the non-smeared Polyakov loop.

With this smeared Polyakov loop field, we construct the squared order parameter
\begin{equation}
\bar{l}_{s}^2 = \dfrac{1}{N_s^3} \sum_{\vec{x}}  |l^{(n)}_{s}(\vec{x})|^2.
\label{eq:op1}
\end{equation} 
However, based on our tests on the lattice, this order parameter still struggles to resolve the critical bubble at large lattice sizes. Inspired by Ref.~\cite{Gould:2024chm}, we construct alternative order parameters designed to further reduce the magnitude of the bulk fluctuations. To motivate these let us write the order parameter schematically 
$l_s(x) = \ell + \delta(x)$, where $\ell$ is the confining phase expectation value of $l_s$ and $\delta$ characterizes the local fluctuation.  Now $l_s(x)^2 \approx \ell^2 + 2\ell\delta(x)$. Applying this to Eq.~(\ref{eq:op1}), the fluctuations of $\bar l_s^{(n)}$ are expected to be first order in volume average of $\delta$.

On the other hand, the fluctuations of the quantity $(l_s(x) - \ell)^2 = l_s(x)^2 - 2\ell l_s(x) + \ell^2 = \delta(x)^2$ are second order in $\delta$, leading us to consider a quantity
\begin{equation}
\bar{l}_{\theta} =  \dfrac{1}{N_s^3} \left(  \sum_{\vec{x}}  |l^{(n)}_s(\vec{x})|^2 - 2A \sum_{\vec{x}} | l^{(n)}_s(\vec{x}) | \right). \label{eq:theta_order_parameter}
\end{equation}
Here constant $A$ can be chosen freely, but in order to optimize the cancellation of fluctuations should be chosen to be close to the confining phase average value of the smeared Polyakov loop. For all our $\ltheta$ runs, shown in Table~\ref{tab:param}, we use $A=0.05$.

Additionally, we test a measurable which we derive following the above argumentation but where $\ell$ is the instantaneous volume average of $l_s$ instead of the expectation value, i.e. $\ell$ varies configuration by configuration. This gives us an order parameter
\begin{equation}
\bar{l}_{\sigma} =  \dfrac{1}{N_s^3}  \sum_{\vec{x}}  |l^{(n)}_s(\vec{x})|^2 - \left( \dfrac{1}{N_s^3}\sum_{\vec{x}} | l^{(n)}_s(\vec{x}) | \right) ^2. \label{eq:sigma_order_parameter}
\end{equation}
This quantity is clearly related to the susceptibility of the smeared Polyakov loop.
It cannot differentiate between the confined and deconfined phases, but rather differentiates between homogeneous bulk phase and mixed-phase configurations. However, since our simulations are generally restricted to only the confinement phase peak and the mixed-phase bubble configurations, this measurable can be used as a pseudo-order parameter.

The same order parameter must be used for generating the multicanonical weight function, and for the multicanonical accept-reject step; the weight functions for different order parameters are not in general similar and need to be determined separately for each.
In the main data collection runs, where the multicanonical weight is not iterated, we measure the distribution of the selected order parameter, as well as the distributions of the other potential order parameters not used for the multicanonical weight function.

As a final note, the value of the action itself could be used to distinguish the phases, but it is very noisy. Often the confinement and deconfinement bulk phase peaks overlap slightly, so resolving the bubble branch would be quite difficult. Unfortunately, however, reweighting the measured order parameter distribution to different $\beta$ is more restricted with a Polyakov loop-based order parameter than it would be with the action \cite{Ferrenberg:1988yz, Moore:2001vf}.

\begin{figure*}[tb]
   \centering    
       \includegraphics[width=0.05\textwidth]{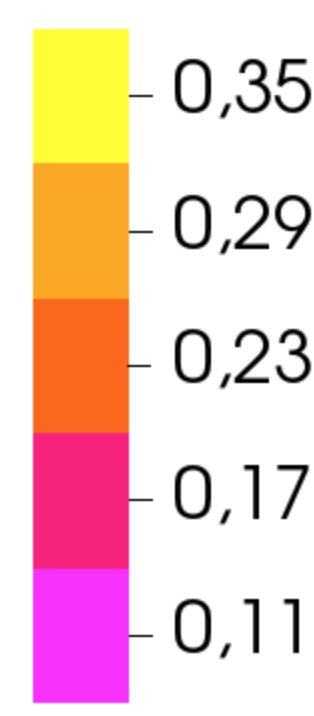} 
   \includegraphics[width=0.3\textwidth]{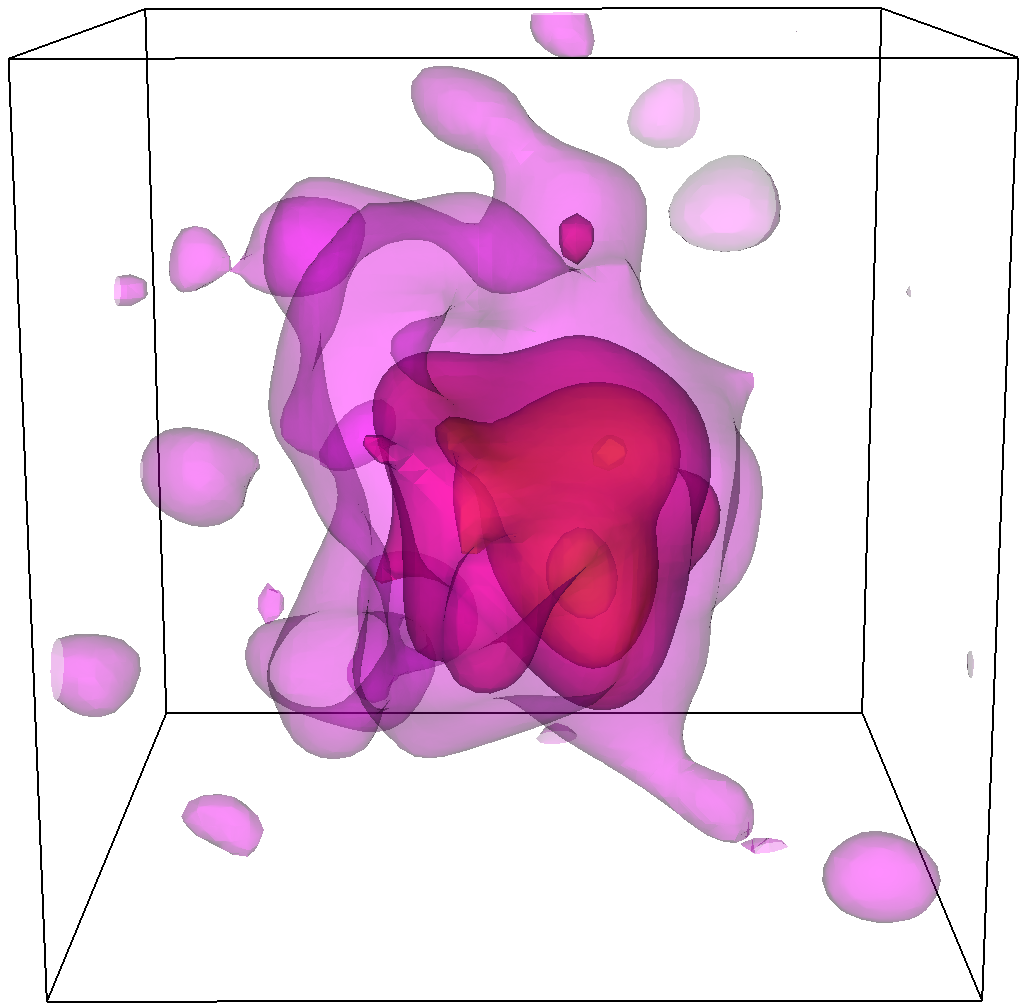}   \includegraphics[width=0.3\textwidth]{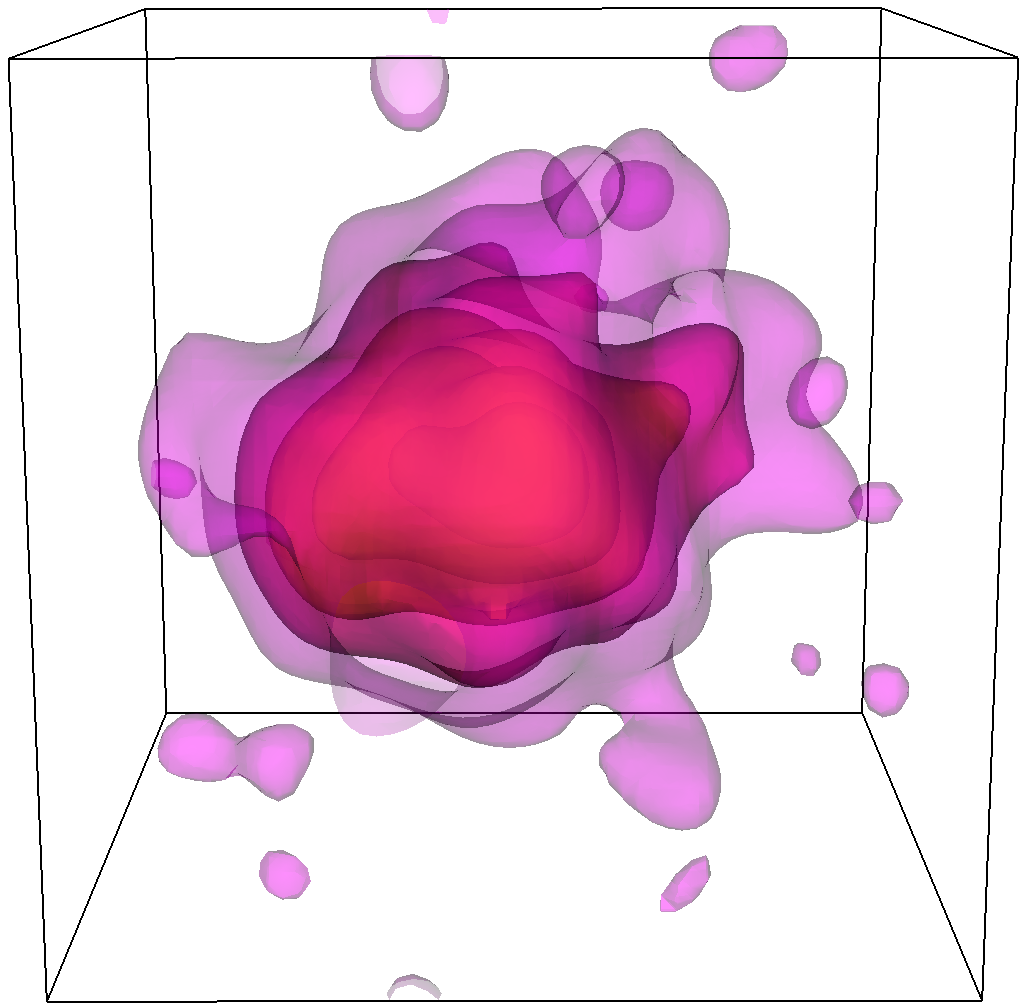} 
   \includegraphics[width=0.3\textwidth]{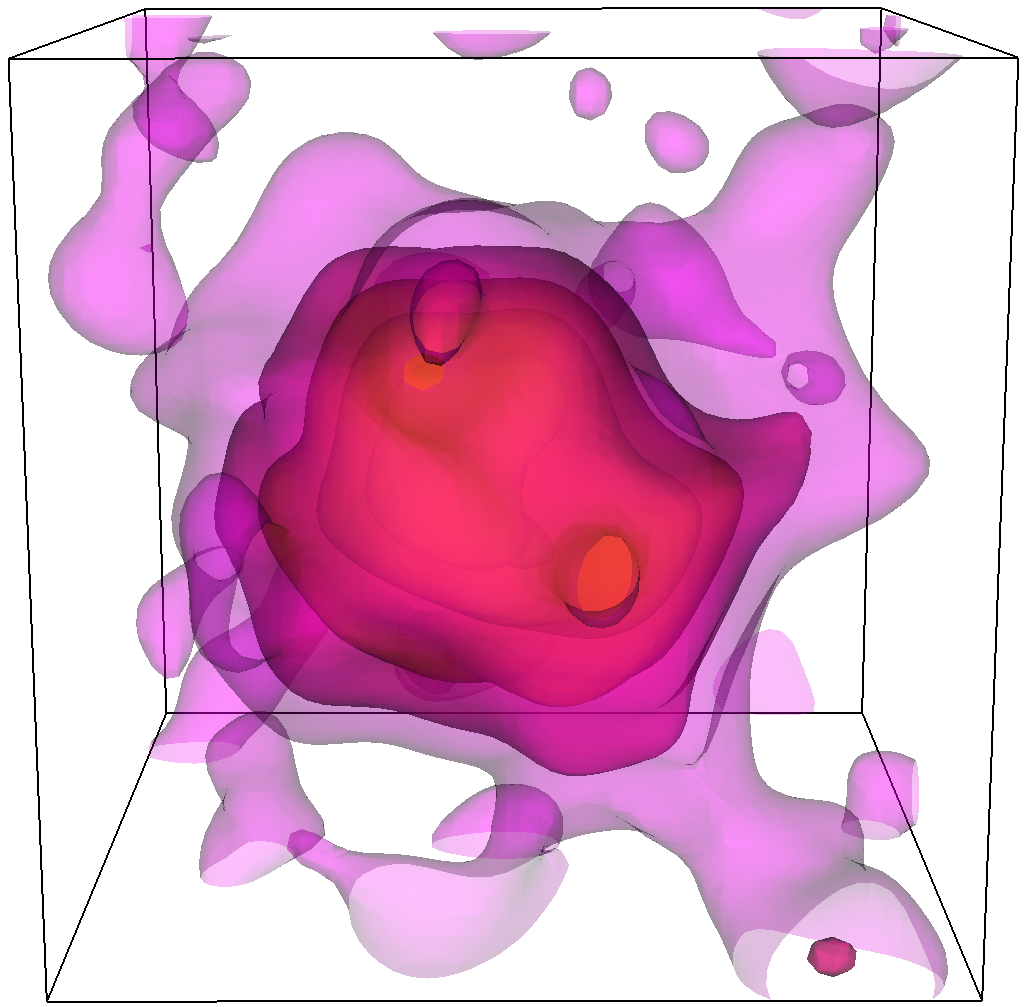}

   \caption{Set of three example configurations from a $\beta = 44.742, N_s = 60$ run where the order parameter was $\mathcal{O} = \ltheta $ from Eq.~(\ref{eq:theta_order_parameter}). Shown are the isosurfaces of the absolute value of smeared Polyakov loop $l_{s}^{(48)}$, as defined in Eq.~(\ref{eq:smeared_loop_field}) , recorded at each spatial site. Note that the configurations appear three dimensional even though our simulation itself is four dimensional, because we are plotting a Polyakov loop-based order parameter. For reference, the deconfinement phase peak value is around 0.41. Configurations are numbered from left to right as 1, 2 and 3, for comparison with Figures~\ref{fig:three_histo} and~\ref{fig:2dhist}.}
   \label{fig:configs}
\end{figure*}

\begin{figure*}[tb]
   \centering \includegraphics[width=\textwidth]{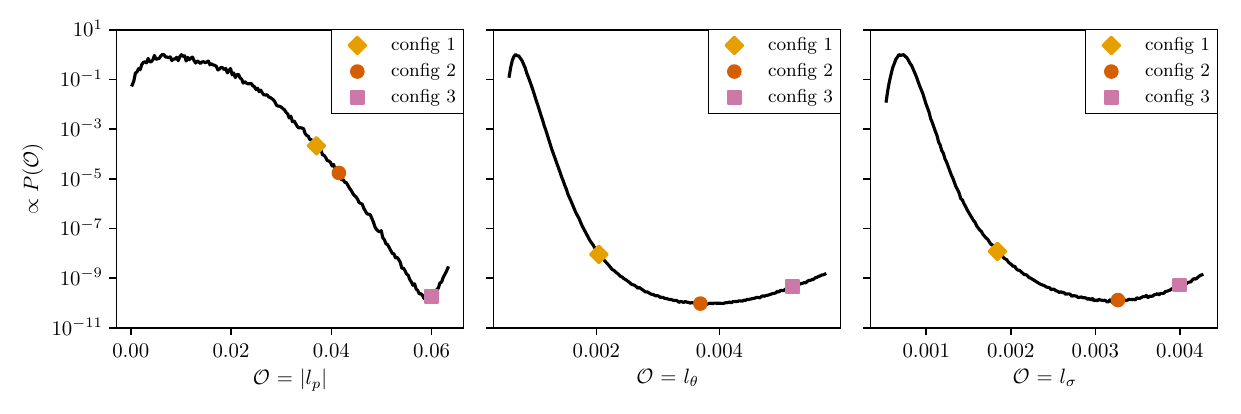} 
   \caption{Histograms for $\beta = 44.742, N_s = 60$, where the order parameter $\mathcal{O}$ used in the multicanonical weighting 
   was, from left to right, $|l_p|$, $\ltheta$ or $\lsigma$, as defined in Eqs.~(\ref{eq:old_order_parameter}), (\ref{eq:theta_order_parameter}) and (\ref{eq:sigma_order_parameter}). Order parameter values corresponding to the configurations shown in Figure~\ref{fig:configs} are marked. Histograms are normalized so that the metastable bulk peak height is one; for our purposes only the difference between the maximum and minimum of the histogram matters. The Gaussian bulk phase peak is significantly wider for $\mathcal{O} = |l_p|$ than for the other two choices.}
   \label{fig:three_histo}
\end{figure*}

\section{Results} \label{sec:results}

We will first compare the volume average of the Polyakov loop, $|\bar{l}_p|$, defined as in Eq.~(\ref{eq:old_order_parameter}), with our two alternative order parameters, $\bar{l}_{\theta}$ and $\bar{l}_{\sigma}$, defined respectively in Eqs.~(\ref{eq:theta_order_parameter}) and~(\ref{eq:sigma_order_parameter}).~Going forwards, we shall drop the bars and thus all mentions of these order parameters can be assumed to be volume averages unless otherwise stated. Furthermore, we will use $\mathcal{O}$ to refer to the order parameter which is used for the multicanonical weight generation and accept-reject steps in a particular run.

\subsection{Order parameter comparison}

To demonstrate the differences between the order parameters, in Fig.~\ref{fig:configs} we show three bubble-like example configurations recorded from a run with $\mathcal{O} = \ltheta$, as defined in Eq.~(\ref{eq:theta_order_parameter}), and parameters $\beta = 44.742$, $N_s = 60$. The configurations are plotted as isosurfaces of the smeared Polyakov loop field $l_{s}^{(48)}$ recorded separately at each spatial site. For each of the example configurations in Fig.~\ref{fig:configs} we record the
corresponding order parameter values $|l_p|$, $\ltheta$ and $\lsigma$.

We can then examine where these configurations would lie in terms of the histograms for different choices of $\mathcal{O}$.
For this purpose we run equivalent simulations with $\mathcal{O} = |l_p|$ and $\mathcal{O} = \lsigma$, and show the histograms of the three different order parameters in Fig.~\ref{fig:three_histo}.
The range of the sampling and thus the histograms are restricted to the confined bulk phase peak and mixed configurations up to bubbles larger than the critical bubble. 

The left-most histogram in Fig.~\ref{fig:three_histo} shows the distribution of the unimproved $|l_p|$, Eq.~(\ref{eq:old_order_parameter}), when we take $\mathcal{O} = |l_p|$ as well. The values $|l_p|$ gets when calculated from the three example configurations from Fig.~\ref{fig:configs} are marked on the histogram. Two of the droplet-like configurations fall under the bulk phase peak, which indicates that $|l_p|$ cannot distinguish between
bulk configurations and droplet configurations at this lattice volume. 

The middle histogram shows the $\ltheta$ distribution, Eq.~(\ref{eq:theta_order_parameter}), when we take $\mathcal{O} = \ltheta$. Finally, the rightmost shows $\lsigma$, Eq.~(\ref{eq:sigma_order_parameter}), when we take $\mathcal{O} = \lsigma$, in line with the other two histograms. For these latter two, the example droplet configurations are well separated from the bulk phase peak, indicating that both $\ltheta$ and $\lsigma$ resolve the bubble configurations well.

Comparing three configurations offers only qualitative evidence for the quality of the order parameters, but it does reveal that $|l_p|$ is not able to identify the bubble configurations accurately, and that the improved order parameters do a better job. This can also be inferred directly from the histogram shapes in Fig.~\ref{fig:three_histo} by comparing to the histogram shape the thin wall approximation predicts: in the histogram of $|l_p|$, the bulk peak covers almost the whole droplet range and there is a sharp transition to droplet/cylinder branch. On the other hand, the two improved order parameters are able to filter out the bulk fluctuations sufficiently to resolve the different branches sketched in Fig.~\ref{fig:thinw}, in particular the droplet branch.

\begin{figure}[tb]
   \includegraphics[width=\linewidth]{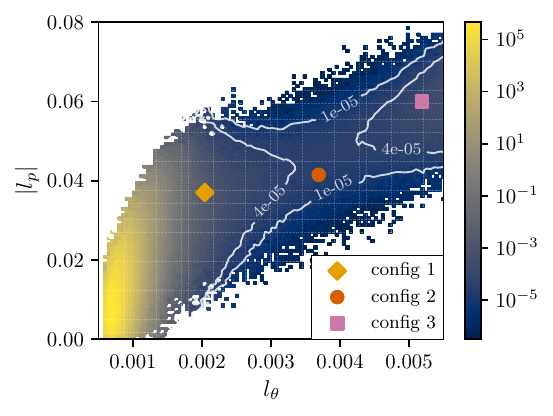} 
   \caption{Two dimensional histogram from $\beta = 44.742, N_s = 60, \mathcal{O} = \ltheta$ runs, with the order parameter $\ltheta$ shown together with the non-smeared Polyakov loop volume average $|l_p|$. Again, the parameter values corresponding to the configurations shown in Figure~\ref{fig:configs} are marked. One value of $|l_p|$ can be seen to correspond to a wide range of $\ltheta$ values both in the bulk phase and the mixed phase. The mixed phase branch can be seen at an angle to the bulk phase branch. Isocontour lines of relative bin frequency are shown to guide the eye. The contours are obtained after one nearest-neighbour averaging of the bin weights. Note that the two observables, $\ltheta$ and $|l_p|$, are correlated.}
   \label{fig:2dhist}
\end{figure}

To further illustrate how $|l_p|$ cannot resolve the droplet branch, we show a 2D histogram of $\ltheta$ and $|l_p|$, from a run with $\mathcal{O} = \ltheta$, in Figure~\ref{fig:2dhist}. Here one can see that one value in $|l_p|$ can correspond to an extremely wide range of values of $\ltheta$ in the range of interest.

It should again be noted that it is volume dependent how well an order parameter identifies the bubbles, but on our range of lattice sizes, $l_{\theta}$ and $l_{\sigma}$ resolve the bubble configurations almost equally well. In the following analysis and measurement of the critical bubble free energy, we will therefore mainly use $\ltheta$, as it is closer to a traditional order parameter. Nevertheless, results of the critical bubble probability for 
both are listed in Table~\ref{tab:results}.

\begin{table}[tb]
\begin{center}
\begin{ruledtabular}
\begin{tabular}{c|c|c|c|c|c}
$\beta$ & $\mathcal{O}$ & $n_{\text{smear}}$ & $N_s$ & $n_{\text{tunnel}}$ & $-\log P_c$ \\
\hline
44.682 & $\ltheta$ & 48 & 72 & 14 & 55.10(31)\\ 
& & & 80 & 17 & 54.05(47)\\
\hline
44.712 & $\ltheta$ & 48 & 60 & 47 & 33.52(27)\\ 
& & & 64 & 33 & 33.99(36)\\ 
& & & 80 & 27 & 34.22(30)\\ 
 & $\lsigma$ & 48 & 60 & 21 & 33.09(35)\\
\hline
44.742 & $\ltheta$ & 48 & 60 & 92 & 23.15(18)\\ 
& & & 64 & 79 & 22.31(20)\\ 
& & & 72 & 51 & 22.63(24)\\ 
& & & 80 & 77 & 22.18(20)\\ 

 & $\ltheta$ & 24 & 60 & 95 & 21.91(17)\\
 & $\lsigma$ & 48 & 60 & 43 & 23.06(26)\\ 
& & & 64 & 44 & 22.27(24)\\ 
& & & 72 & 27 & 22.92(36)\\

\end{tabular}

\end{ruledtabular}

\end{center}
\caption{Results of the critical bubble free energy $F_c/T = -\log P_c$ measured with the difference in histogram minimum and maximum, as per Eq.~(\ref{eq:Pc_approx}). The indicated order parameter $\mathcal{O}$ is used both for the multicanonical updates and for the critical bubble probability calculation. Also shown is $n_{\text{tunnel}}$, the number of histogram traversals from the bulk phase peak to the critical bubble order parameter value, or back. Errors are obtained by bootstrap resampling, and are purely statistical. }
\label{tab:results}
\end{table}

The level of smearing was also briefly investigated by using 
$\mathcal{O}=l_{\theta}$ with half the number of smearing steps we originally used (i.e. 24 instead of 48 steps).  The exact level of smearing should not be important for this application as long as we are not smearing too much or too little. Too little smearing degrades the efficiency of the order parameter at identifying the bubbles, whereas with too much smearing we start to lose the bubbles. This does not happen if the smearing radius is smaller than the size of the critical bubble. For all of the parameters we simulate, the smearing radius, which is proportional to the square root of the number of smearing steps \cite{Rindlisbacher:2025dqw}, is well below the critical bubble radius. With this consideration, and since the two levels of smearing give comparable results (see Table~\ref{tab:results}), we will use the 48-step smearing for our main results.

\begin{figure}[tb]
   \centering \includegraphics[width=0.9\linewidth]{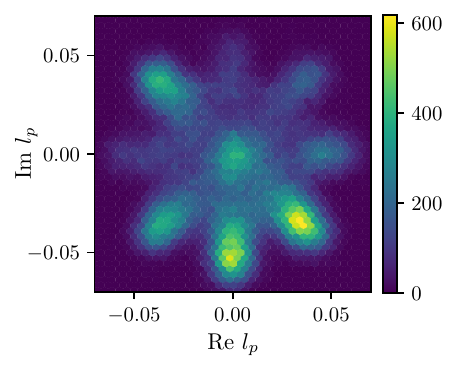} 
   \caption{Sampled values of $l_p$ 
   from a simulation with $\mathcal{O} = l_{\theta}$,
   $N_s = 60$, $\beta = 44.742$. Note that this shows the non-reweighted distribution of $l_p$. The range of $|l_p|$ is restricted to the confined and mixed phases as usual: only the metastable bulk phase peak and mixed-phase configurations up to volume fractions slightly bigger than critical bubble are sampled.}
   \label{fig:branches}
\end{figure}

As a final note, we expect the alternative order parameters allow the system to tunnel to any of the eight degenerate confined vacua equally likely. We confirm that this is the case for all of our alternative order parameters. To illustrate this, in Fig.~\ref{fig:branches} the sampled values of the imaginary and real parts of the Polyakov loop $l_p$ are shown for simulation realization with $\mathcal{O} = l_{\theta}$, $N_s = 60$, $\beta = 44.742$. The plotted distribution is not reweighted: the purpose of the figure is only to show the typical sampled values of $l_p$, which reveal the 8-branched structure of the degenerate deconfinement branches.

\subsection{Are these bubbles really critical?}

The usual recipe for calculating the nucleation rate 
on the lattice involves evolving critical bubbles in (real) time to measure how often they lead to tunneling from one phase to another. This is done by evolving a critical bubble configuration forwards and backwards in time. If the trajectories end up in different phases, the initial critical configuration led to tunneling.

As we are in a strongly coupled regime, this method is currently unavailable to us. However, we would like to be able to say with some certainty that the bubble configurations corresponding to the order parameter values at the minimum of the order parameter distribution are actually critical in the sense that they correspond to a saddle-point in free energy landscape.

With this aim in mind, we mimic real time evolution by performing only heat bath updates without multicanonical sampling. This is inspired by the relationship between Langevin time and heat bath steps which exists for weak coupling~\cite{Bodeker:1998hm}. No such relation has been
found for strong coupling. However, the heat bath evolution serves as a sort of qualitative check: if we see no bubble configuration resulting in tunneling from one phase to the other, we can conclude that our bubbles are likely not critical.

\begin{figure}[tb]
   \centering \includegraphics[width=\linewidth]{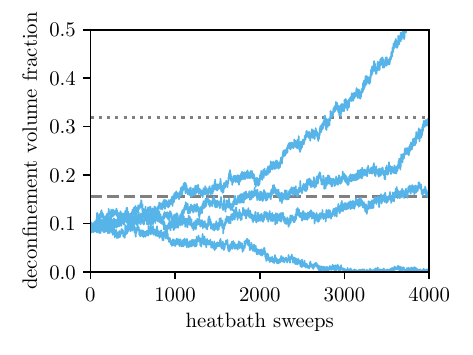} 
   \caption{Example starting configuration with four evolution trajectories, from a $N_s = 60$, $\beta = 44.742$, $\mathcal{O} = \ltheta$ run. The configuration is evolved four times with only heatbath sweeps to see which phase it ends up in: this example configuration leads to tunneling trajectories between the stable and metastable phases. The $y$-axis is the approximate volume fraction taken up by the deconfined phase, estimated from the order parameter value. 
   The dashed line is the theoretical limit for bubble branch stability, while the dotted line is the limit for the cylinder branch. }
   \label{fig:trajectories}
\end{figure}

Our procedure is the following. We save a configuration which has an order parameter value close to the critical value, and separately evolve that configuration four times with only heatbath steps (no multicanonical, no overrelaxation) until it reaches either the metastable (confined) or stable (deconfined) phase. We then glue pairs of these trajectories together, with one of the pair corresponding to `evolution' backwards in time and one to forwards in time. One then sees if any of the glued trajectories show tunneling from one phase to the other.

A typical example of this procedure is presented in Fig.~\ref{fig:trajectories}: our example configuration leads to one trajectory going to confined phase, two trajectories going to the deconfined phase, and one trajectory that is stuck around the order parameter values of the critical bubble. This kind of hanging-around happens often, and is possibly due to the barrier between different mixed-phase topological configurations. The starting configuration does thus lead to some trajectories tunneling between deconfined and confined phases, at least according to this naive analysis. 

As we observe several tunneling trajectories, we conclude that considering the limitations, taking $\mathcal{O}=\ltheta$ does accurately resolve something strongly resembling a critical bubble. We will now go on to present results for the critical bubble probability.

\begin{figure*}[tb]
   \centering
   
\subfloat[Results at $\beta = 44.682$]{
   \includegraphics[height=0.35\linewidth]{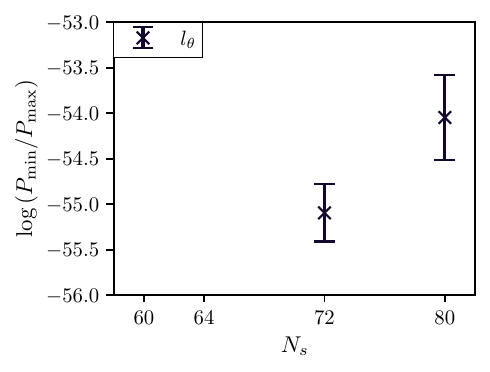} 
   \includegraphics[height=0.35\linewidth]{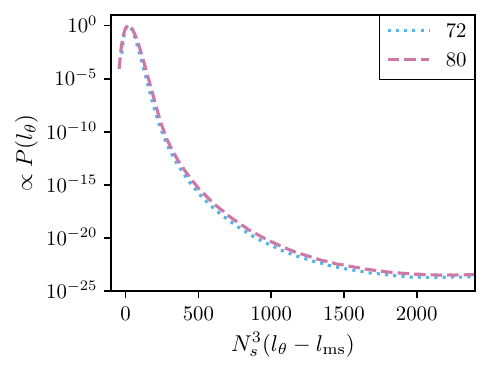} 
}\\
\subfloat[Results at $\beta = 44.712$]{
   \includegraphics[height=0.35\linewidth]{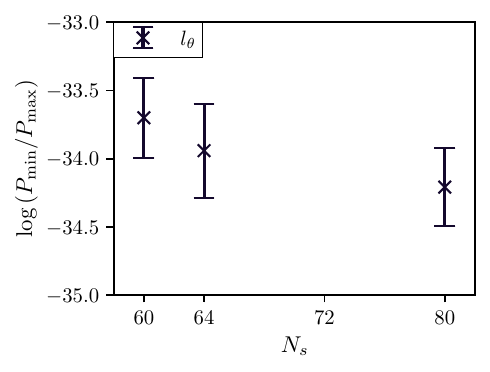} 
   \includegraphics[height=0.35\linewidth]{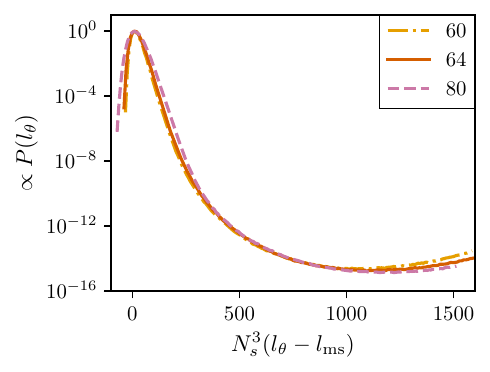}
}\\
\subfloat[Results at $\beta = 44.742$]{
   \includegraphics[height=0.35\linewidth]{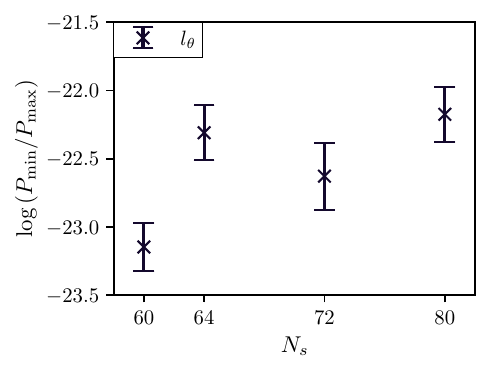} 
   \includegraphics[height=0.35\linewidth]{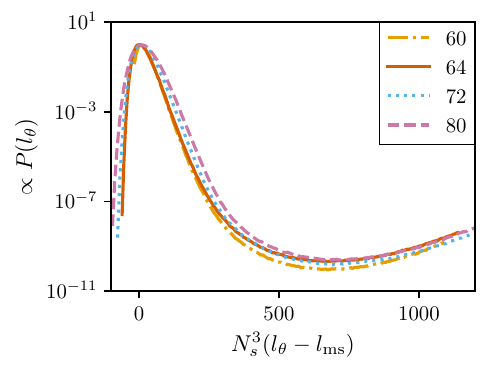} 
}
       
   \caption{The leftmost plots show the critical bubble probability as defined in Eq.~\ref{eq:alternative_Fc} obtained at different lattice sizes. On the right the histograms at different volumes are presented. The $x$-axes of the histograms are normalized with spatial volume and $l_\text{ms}$, the order parameter value corresponding to the metastable phase peak; shown here is thus an extensive order parameter constructed from $\ltheta$. The histogram peak height is normalized to one.}
   \label{fig:volume_histograms}
\end{figure*}

\subsection{Probability of a critical bubble}

The following results are obtained with $\mathcal{O}=l_{\theta}$ as the order parameter used for multicanonical sampling. The critical bubble probabilities obtained from the height difference of the minimum and maximum of the measured order parameter distribution are presented in Fig.~\ref{fig:volume_histograms}, together with the corresponding histograms. Table~\ref{tab:results} lists the results for different sets of parameters we simulated.

Errors are obtained by bootstrap resampling, 
with the number of bootstrap blocks equal to the number of histogram traversals in either direction. One block then contains approximately one histogram traversal, and the block length is much larger than the autocorrelation length. The number of resamplings of the blocks used for errors in Table~\ref{tab:results} is 500. We observe that the errors remain stable with increased and decreased numbers of resamplings. The quoted errors are purely statistical: we do not account for systematic uncertainties that might arise from, for example, the dependence of the result on the chosen order parameter, or from smearing.

We observe little dependence on the spatial lattice volume $V$. 
Let us consider an extensive order parameter. With increasing spatial volume, the Gaussian peak width increases as $\sqrt{V}$, and correspondingly the peak height decreases as $1/\sqrt{V}$. Since we obtain the probability of the critical bubble by comparing the histogram maximum and minimum, there is a $\sqrt V$ increase in the relative probability with increasing volume. 
On the other hand, for a bubble of fixed size, the number of possible bubble locations grows with volume, which means the critical bubble minimum goes up as $V$~\cite{Moore:2000jw}. 

These two effects combined would mean the critical bubble probability increases with $V^{3/2}$ as the volume goes up. Between our minimum volume $60^3$ and maximum volume $80^3$ this corresponds to an increase of a factor of approximately 3.6 in the relative probability; this means the difference in the logarithms of the probabilities should be about 1.3.

This change is comparable to the size of our estimated errors, but the smallest and largest $\beta$ in Fig.~\ref{fig:volume_histograms} would nevertheless seem to indicate this kind of behaviour. The middle $\beta$ shows more peculiar behaviour with seemingly the opposite effect. This might be due to the bubble being too large to fit the $N_s = 60$ lattice properly, so it is `feeling' itself through the walls and becoming deformed. This effect is not visible for the larger $\beta$ at $N_s = 60$, since increased superheating makes the critical bubble smaller.

The $\sqrt V$ scaling effect is due to the use of the maximum-minimum histogram method to obtain the probability, and would disappear if we used the integral method of Eq.~(\ref{eq:integral_Pc}). This however would give us the bubble probability in units of inverse order parameter, and in any case the $V$ scaling coming from translational zero modes of the bubble would still remain. The dynamical prefactor, which we cannot access, would cancel both of these effects exactly, leading to a well defined order parameter-independent nucleation rate. 

It remains unclear to us how to resolve these issues without the dynamical prefactor, and as such the critical bubble probability does not have a well defined infinite volume limit. Therefore we will use the largest volume result ($N_s= 80$) for the probability, obtained with the minimum-maximum histogram method, and compare this to the free energy obtained with thin wall approximation.

Additionally we stress that we simulate only at a single temporal extent.
We therefore do not obtain continuum limit results and can only 
speculate on the effect of varying the lattice spacing. Previous work considering bubble nucleation in scalar field theories would indicate that these can be significant~\cite{Moore:2001vf}.

\section{Comparison to the Thin Wall approximations}

\label{sec:compar}

We will now compare our lattice results to thin wall estimates for the rate. The thin wall approximation simplifies nucleation rate calculations significantly, but its range of applicability is limited. It holds only for small superheating or -cooling and a strongly suppressed rate, when the critical bubble is very large compared to the bubble wall thickness~\cite{Linde:1981zj}. Additionally the thin wall approximation ceases to be useful at large $N_c$, number of colors, though $N_c = 8$ still seems to be small enough for it to work~\cite{Huber:2025qbl}.

While we only simulate at one temporal extent $N_t$ and thus cannot obtain a continuum extrapolated value for the critical bubble probability, we can compare our results to thin wall estimates. In particular, as the thin wall calculations can be performed using lattice measurements of surface tension and latent heat, we can compare to estimates obtained with continuum-extrapolated values, 
as well as with values obtained at our choice of temporal extent $N_t = 6$. 

\begin{figure*}[tb]
   \centering \includegraphics[width=1\linewidth]{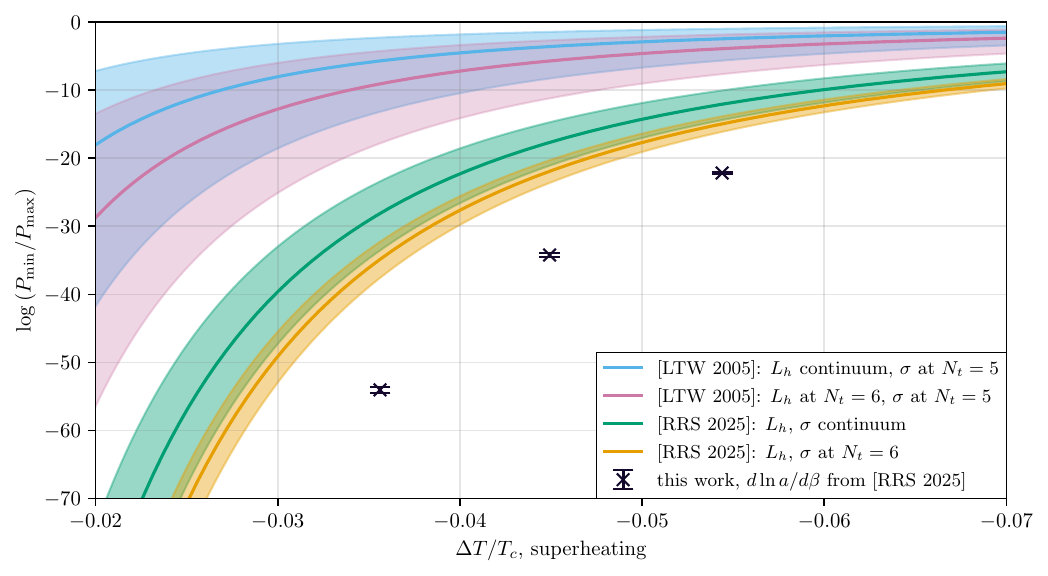} 
   \caption{Comparison of the free energy of a critical bubble obtained with the thin wall method, and the critical bubble probability obtained with our simulation method. The highest and second-highest bands correspond to thin wall results obtained using Ref.~\cite{Lucini:2005vg} (denoted in the caption as LTW~2005) with either the continuum value of the latent heat 
and the value of latent heat at $N_t = 6$, respectively. The lower two bands correspond to thin wall results obtained using values from Ref.~\cite{Rindlisbacher:2025dqw} (denoted in the caption as RRS~2025); the higher one with continuum results for latent heat and surface tension, and the lower with values at $N_t = 6$. The scatter points correspond to our 
results for $N_s = 80$, $N_t = 6$ and using order parameter $\mathcal{O}=\ltheta$. Superheating in $\Delta \beta$ is translated to superheating in $\Delta T/T_c$ using the lattice beta function from Ref.~\cite{Rindlisbacher:2025dqw}. Errors for the scatter points are obtained by bootstrapping the measurements of the order parameter, while error bands on the thin wall estimates 
are based on the reported errors in Refs.~\cite{Lucini:2005vg} and~\cite{Rindlisbacher:2025dqw}. }
   \label{fig:thin_wall-probs}
\end{figure*}

In the thin wall approximation, the free energy of a droplet of radius $r$ at temperature $\Delta T = T_c - T$ is given by
\begin{equation}
F(r) = 4 \pi r^2 \sigma - \frac{4\pi}{3} r^3 L_h \dfrac{\Delta T}{T}.
\end{equation}
The first term is the contribution due to surface tension $\sigma$, which is proportional to the surface area of the droplet, while the second term is the contribution from latent heat $L_h$, which is proportional to the droplet volume.

The droplet radius which maximises the free energy is termed `critical'. This happens at radius and free energy
\begin{align}
r_c &= \dfrac{2\sigma}{L_h} \left(\frac{\Delta T}{T}\right)^{-1},\\ 
F_c &= \frac{16\pi}{3} \dfrac{\sigma^3}{L_h^2} \left( \frac{\Delta T}{T}\right)^{-2}.
\end{align}

As usual in the thin wall approximation, we take the surface tension and latent heat to be independent of temperature $T$. This should hold fairly well close to the transition, where the method is expected to be more accurate. Latent heat and surface tension are taken as inputs, with lattice results for $L_h/T_c^4 = \tilde{L_h}$ and $\sigma/T_c^3 = \tilde{\sigma}$ as well as the lattice beta function from Refs.~\cite{Lucini:2005vg} and~\cite{Rindlisbacher:2025dqw}.

With these we write the critical bubble free energy $F_c$ as 
\begin{equation}
\dfrac{F_c}{T} = \frac{16\pi}{3} \dfrac{\tilde{\sigma}^3}{\tilde{L_h}^2} \dfrac{T_c}{T} \left( \frac{\Delta T}{T}\right)^{-2} .
\end{equation}

Approximating $T \sim T_c$, we get
\begin{equation}
\dfrac{F_c}{T} = \frac{16\pi}{3} \dfrac{\tilde{\sigma}^3}{\tilde{L_h}^2} \left( \frac{\Delta T}{T_c}\right)^{-2} . \label{eq:thin_wall_F}
\end{equation}

The lattice beta function $d\ln(a)/d\beta$ is needed to relate the lattice coupling $\beta$ to the temperature $T$.
As $T = 1/(N_t a)$, we have
\begin{equation}
\frac{d \ln(a)}{d\beta} = - \frac{1}{T} \frac{dT}{d\beta},
\end{equation}
which with some manipulation gives
\begin{equation}
\dfrac{\Delta T}{T_c} = 1 - \dfrac{1}{1 - \Delta \beta \dfrac{d\ln(a)}{d\beta} }.
\end{equation}

In Figure~\ref{fig:thin_wall-probs} we plot the estimated $-F_c/T$ obtained with Eq.~(\ref{eq:thin_wall_F}) against our results for $\log(P_c)$ at $N_s = 80, N_t=6$. For the thin wall estimates, we include both the continuum result and the $N_t = 6$ results.

In Ref.~\cite{Seppa:2025lud} we reported that the thin wall estimate for the rate is off from our result by many orders of magnitude. While this is the case for the results obtained in Ref.~\cite{Lucini:2005vg}, the newer measurements of the latent heat, surface tension, and lattice beta function presented in Ref.~\cite{Rindlisbacher:2025dqw} show better agreement with our result. Our result indicates the lattice estimate of the critical bubble probability is lower by a factor of $e^{-5}$ to $e^{-10}$. Earlier works in the cubic anisotropy model~\cite{Moore:2001vf} and scalar field theory~\cite{Gould:2024chm} show similar increase in suppression in the lattice results compared to thin wall calculations.

We also note that while the thin wall result and our result still differ, it is plausible that if proper continuum extrapolation were possible our result would move closer to the thin wall result. However, for this reason we compare also to the thin wall calculation obtained using lattice results from $N_t = 6$. The discrepancy between the thin wall calculation and the critical bubble probability obtained from the lattice persists there as well.

As it stands, however, the critical bubble probability we obtain
from the lattice is indeed order parameter dependent, and the proper way to take the infinite volume limit is unclear if we are not computing a full nucleation rate.
Instead, we have naively compared the thin wall calculation directly to our result obtained on the largest lattice we use.

\section{Discussion} \label{sec:disc}

In this paper, we have carried out the first direct lattice measurements of the critical bubble free energy in a confining model. Our aim was to obtain an estimate for the nucleation rate in SU(8) deconfinement transition, and to compare this to the corresponding thin wall estimate. 

We obtain a relative critical bubble free energy, which gives approximately the statistical part of the nucleation rate, for three different levels of superheating. The result is obtained only for one temporal extent $N_t = 6$. To compute the result, we need to resolve the critical bubbles accurately, meaning the order parameter needs to differentiate between bubble configurations, other mixed phase configurations, and bulk phase configurations. With a smeared Polyakov loop based order parameter we are able to accurately resolve the critical bubble. This result highlights the importance of choosing a suitable order parameter in these types of calculations.

If the nucleation rate is estimated directly as $\exp(-F_c/T)$, our result gives a rate suppressed by between $e^{-5}$ to $e^{-10}$ compared to the thin wall estimate obtained with the newest lattice data for the latent heat and surface tension~\cite{Rindlisbacher:2025dqw}. Thin wall approximation is not expected to be very accurate further away from the transition, so this is not unexpected. However, we only obtain the result at one temporal extent, and proper continuum extrapolation might bridge some of the gap between the results. The computation is also order parameter dependent, although we obtain very similar results with both our alternative order parameters. This might indicate that there is no strong dependence on the order parameter if it resolves the critical configurations accurately enough. However, since the two improved order parameters share many similarities by construction, it is hard to draw clear conclusions.

It should also be noted that our result is valid only for the deconfinement transition. It is not a priori clear that the confinement direction of the transition would have the same critical bubble free energy. Indeed at least the maximum supercooling is expected to be reduced compared to the maximum superheating~\cite{Agrawal:2025xul}. This is an interesting topic for further study.

The present method does not accurately give the whole nucleation rate, since we are unable to analyse the dynamics of bubble evolution. This is another area for further study; possibly the method of Ref.~\cite{BarrosoMancha:2022mbj} could be extended to obtain at least some dynamical information on the rate. Nevertheless, this work presents a step towards determining the nucleation rate nonperturbatively in strongly coupled models.

\begin{acknowledgments}
The simulations for this paper were carried out at the
Finnish Centre for Scientific Computing CSC. 
We wish to thank Oliver Gould for useful discussions and Tobias Rindlisbacher for detailed comments on the manuscript. RS was supported by the Magnus Ehrnrooth Foundation, and Research Council of Finland grant 349865, and the European Research Council grant \textit{CoCoS}, no.~101142449.
DJW was supported by Research Council of Finland grant nos. 324882, 349865 and 353131.
KR was supported by the European Research Council grant \textit{CoCoS}, no.~101142449, and the
Research Council of Finland grant no.~354572.
\end{acknowledgments}

\section*{Data access statement}

The simulation data is openly available at \url{https://doi.org/10.5281/zenodo.19184044}.  

\bibliography{references}

\clearpage

\end{document}